\documentclass{PoS}

\usepackage{subfigure}
\PoS{PoS(HEP2005)249}

\title{Final state interactions and long-distance effects \\in $B\to \pi\pi K$ and 
$B\to K\bar KK$ decays \thanks{This work has been performed in the framework of the IN2P3-Polish
laboratories Convention (project No. 99-97).}}

\author{\speaker{Leonard Le\'sniak$^{a}$}, Agnieszka Furman$^a$, Robert
 Kami\'nski$^a$ and Benoit Loiseau$^b$\\
$^{ a}$  Department of Theoretical Physics, H. Niewodnicza\'nski Institute
 of Nuclear Physics,\\\hspace{0.2cm} Polish Academy of Sciences, 31-342 
 Krak\'ow, Poland\\
$^b$ Laboratoire de Physique Nucl\'eaire et de Hautes \'Energies
\thanks{Unit\'e de Recherche des Universit\'es
Paris 6 et Paris 7, associ\'ee au CNRS} , Groupe Th\'eorie,
\hspace{0.2cm}Univ. P. \& M. Curie, 4 Pl. Jussieu, F-75252 Paris, France \\

        E-mail: \email{Leonard.Lesniak@ifj.edu.pl}}

\abstract{Charged and neutral $B$ decays into the $\pi^+\pi^-K$, $K^+K^-K$ and
$K^0_SK^0_SK^0_S$ systems have been studied. The $\pi^+\pi^-$ and $K^+K^-$ final
state interactions in the $S$-wave isoscalar state are treated in an unitary
way. The long-distance contributions called charming penguins are 
introduced in the model. Effective mass distributions, branching 
ratios and some asymmetries are succesfully compared with the recent BaBar and 
Belle data. A particularly large negative direct CP-violating asymmetry 
in the charged $B$ decays into $f_0(980)K$ is evaluated for one set of the 
charming penguin amplitudes.}

\FullConference{International Europhysics Conference on High Energy Physics\\
		 July 21st - 27th 2005\\
		 Lisboa, Portugal}
\ShortTitle{Final state interactions and long-distance effects in $B$ decays}

\begin{document}

\section{Introduction} A good description of final state interactions is 
crucial to
obtain a precise determination of the Cabibbo-Kobayashi-Maskawa matrix 
elements. Recent experimental results on the $B$ decays into three
pseudoscalar mesons are a rich source of information about weak decay
amplitudes and two body mesonic strong interactions.  
Direct CP-violation effects can appear in these reactions. Here we study the
$B$ decays into $\pi\pi K$ and $K\bar KK$ channels. An important part of 
these processes is a production of the $\pi\pi$ or $K\bar K$ 
pairs in the isospin zero $S$-wave. A very prominent maximum of 
the $f_0(980)$ resonance has been seen in the $\pi\pi$ spectra obtained by the
BaBar and Belle collaborations (see, for example refs.[1-4]).
One can expect that the other scalar resonance $f_0(600)$ plays some role
at lower $\pi\pi$ effective masses. Both resonances are incorporated in a
natural way in the unitary three-channel model of pion-pion, kaon-kaon and 
four-pion interactions \cite{kami97}. We use this model to calculate the final 
state interactions in the different $B$-decay channels.

The penguin diagrams are essential in studies of $B$ decays into 
$\pi\pi K$ and $K\bar K K$. However, different penguin
amplitudes interfere destructively in the decay $B\to f_0(980)K$. This leads
to much too small branching ratios calculated in the factorization approach. 
Thus we consider the long-distance 
contributions originating from enhanced charm quark loops. These so-called
charming penguin terms correspond, for example, to weak decays of $B$ to 
intermediate $D_s^{(*)}D^{(*)}$ states followed by transitions to the $f_0(980)K$ 
final states via $c\bar c$ annihilations. The charming penguins have been
used in fits of experimental branching fractions for two-body charmless $B$ 
decays [6-7].
 Below we show that these long-distance contributions
are also needed in analyses of $B$ decays into three pseudoscalar
mesons $\pi\pi K$ or $K\bar K K$.

\section{Model for the \boldmath{$B\to\pi\pi K$} and \boldmath{$B\to 
K\bar KK$ decays}} 
We have constructed a model for the following charged and neutral $B$ decays:
 $B^\pm\!\to\!(\pi\pi)_SK^\pm$, $
B^\pm\to(K\bar K)_SK^\pm$, $B^0\!\to\!(\pi\pi)_SK^0,
\ B^0\!\to\!(K\bar K)_SK^0$, 
$\bar B^0\!\to\!(\pi\pi)_S\bar K^0$ and 
$\bar B^0\!\to\!(K\bar K)_S\bar K^0$.
Here by $(\pi\pi)_S$ and $(K\bar K)_S$ we mean $\pi^+\pi^-$ or $\pi^0\pi^0$ 
and $K^+K^-$ or $K^0\bar K^0$ pairs in isospin zero $S$-wave.

 Below we
outline only the main features of the amplitudes described in more detail
in \cite{FKLL}. The decay amplitudes consist of two parts both related to weak
transitions $b \to s \bar n n$ and $b \to s \bar s s$, where $n$ 
denotes $u$ or $d$ quarks. The first part
corresponds to the factorization approximation with some QCD corrections. The
second one is the long-distance amplitude $A_{LD}$ with $c$- or $u$-quark loops.
Its form for the $B^-\!\to\!(\pi^+\pi^-)_S K^-$ decay reads:

\begin{equation}
  A_{LD}   =\frac{G_F}{\sqrt{3}}\chi \left[C(m_{\pi\pi}) \Gamma_1^{~n*}(m_{\pi\pi})+
  C(m_K) \Gamma_1^{~s*}(m_{\pi\pi})\right],
\end{equation}
where $G_F$ is the Fermi coupling constant, $\chi$ is a constant which value
is 
close to 30 GeV$^{-1}$, $m_K$ is the kaon mass, $\Gamma_1^{~n}(m_{\pi\pi})$ and 
$\Gamma_1^{~s}(m_{\pi\pi})$ are the non-strange and strange pion scalar form 
factors depending on the effective pion-pion mass $m_{\pi\pi}$.
The charming penguin contribution is written similarly as in ref. \cite{groo03}:
\begin{equation}
C(m)=\!-\!\left(M_B^2\!-\!m^2\right)\!f_\pi F_\pi\left(V_{ub}V_{us}^* P_u+
V_{tb}V_{ts}^*P_t \right),
\end{equation}
where $f_\pi F_\pi=0.042$ GeV, $P_t$ and $P_u$ are complex parameters multiplied
by the products of the Cabibbo-Kobayashi-Maskawa matrix elements $V$. We have
performed numerical calculations using two sets of parameters fitted
to data on two-body charmless $B$-decays. We consider the model I
in which \cite{groo03}
$P_t=(0.068\pm 0.007)\exp[i(1.32\pm 0.10)]$ and 
$P_u=(0.32\pm 0.14)\exp[i(1.0\pm 0.27)]$, and the model II \cite{ciuc04} with 
$P_t =(0.08\pm 0.02)\exp[-i(0.6\pm 0.5)]$ and $P_u=0$.

In our treatment of final state interactions we use not only the pion scalar
form factors $\Gamma_1^{~n}(m_{\pi\pi})$ and $\Gamma_1^{~s}(m_{\pi\pi})$ but
also the kaon scalar form factors $\Gamma_2^{~n}(m_{K\bar K})$ and
 $\Gamma_2^{~s}(m_{K\bar K})$. The first and the second form factors are 
responsible for the transitions from the $\bar n n$ quark pair
and from the $\bar s s$  quark pair to the $\pi\pi$ pair,
respectively. Similarly the third and the fourth form factors correspond
to the $\bar n n \to K\bar K$ and to the 
$\bar s s \to K\bar K$ transitions. In the $B$-decay processes
transitions from the $\pi\pi$ channel to the $K\bar K$ channel 
and vice versa are possible. They are incorporated in the two-body scattering 
matrix $T$. The interchannel couplings are included in the following 
formulae:

\begin{equation}
\label{eq:cont}
\Gamma_i^{~n,s}(m)=R_i^{n,s}(m)
+\sum_{j=1}^2\langle k_i \vert R_j^{n,s}(m)G_j(m)T_{ij}(m) \vert k_j \rangle,
\end{equation}
where $\vert k_i \rangle$ and $\vert k_j \rangle$ represent 
the wave functions of the two mesons in the momentum space and the indices 
$i,j=1,2$ refer to the $\pi\pi$ and $K\bar K$ channels, respectively.
The center of mass channel momenta are given by 
$k_1=\sqrt{m^2/4-m_\pi^2}$ and $k_2=\sqrt{m^2/4-m_K^2}$.

In the numerical calculations we use the solution $A$ of the 
coupled channel model~\cite{kami97}.
The functions $G_i(m)$ are the free particle Green's functions defined 
in~\cite{kami97} and $R_i^{n,s}(m)$ are the production functions 
responsible for the initial 
formation of the meson pairs prior to rescattering.
The production functions have been derived by Mei\ss ner 
and Oller in the one-loop approximation of the chiral perturbation theory
~\cite{meis01}.

Let us remark that in this approach a sum of many Breit-Wigner terms typical for
the so-called isobar model, usually
used in phenomenological fits to Dalitz plots, is
replaced by a set of the coupled meson-meson amplitudes $T$.
These amplitudes can be expressed in terms of the phase shifts  
$\delta_{\pi\pi}$, $\delta_{KK}$ and the inelasticity $\eta$ known from other 
experiments. Thus no arbitrary phases nor relative intensity free parameters 
for different resonances are needed. In addition, both the $T$- and the
\mbox{$\Gamma$-matrices} satisfy unitarity constraints. A generalization of 
Watson's theorem to multi-channel $B$-decays is thus realized. 
\section{Results}
Table 1 shows that the results of our calculations compare reasonably well
with the results of the experimental analyses averaged by HFAG in spring 2005.

 \begin{table*}[t]
\begin{center}
\caption{Branching ratios $\mathcal B$ in units of $10^{-6}$, direct CP asymmetry
 $\mathcal A_{CP}$, time
dependent asymmetry parameters $\mathcal A$ and $\mathcal S$ for the  
$B\!\to\!f_0(980)K, f_0(980)\!\to\!\pi^+\pi^-$ decay. The model errors come from
uncertainties of the charming penguin amplitudes.}
\medskip
\begin{tabular}[]{|c|c|c|c|c|}
\hline
	decay mode   &     observable       & experiment                       &  Model I       & Model II\\

 &  &  HFAG  &$\chi=33.5$ GeV$^{-1}$& $\chi=23.5$ GeV$^{-1}$\\
\hline
 \raisebox{-1.5ex}[0pt][0pt]{$B^\pm$}    &$\mathcal{B}$   & $8.49^{+1.35}_{-1.26}$  & $8.49$ (fit)  & $8.46$ (fit) \\
                                          &$\mathcal{A}_{CP}$& $-0.13^{+0.19}_{-0.12}$ &$-0.52\pm 0.12$ & $0.20\pm 0.20$\\
\hline
                                          &$\mathcal{B}$     & $6.0\pm{1.6}$ &$5.9\pm 1.6$    & $5.8\pm 2.8$\\
$B^0$ &$\mathcal{A}$     & $-0.14\pm0.22$&$0.01\pm 0.10$  & $0.0004\pm 0.0010$\\
                                          &$\mathcal{S}$     & $-0.39\pm0.26$&$-0.63\pm 0.09$ & $-0.77\pm 0.0004$\\
\hline

\end{tabular}
\end{center}
\end{table*}
In Fig.1a we compare the model I with the BaBar 
data \cite{baba03} for the $B^\pm\to \pi^+\pi^-K^\pm$ decays. One can see
that the theoretical solid line describes well the shape of the $f_0(980)$ 
resonance. Fig. 1b shows
the Belle data \cite{chen04} in comparison with our model for the neutral
$B^0$ decays to  $\pi^+\pi^-K^0$ in a wide range of effective masses. 
The model provides us with an absolute prediction for the branching ratio
of the $B^0$ decay into $f_0(980)K^0$ if the effective mass 
distributions in the range of the $f_0(980)$ resonance for the charged $B$ decays
are reproduced.

 For the model I one obtains a particularly large negative direct CP-violation
 asymmetry. 

\begin{figure}[!ht]
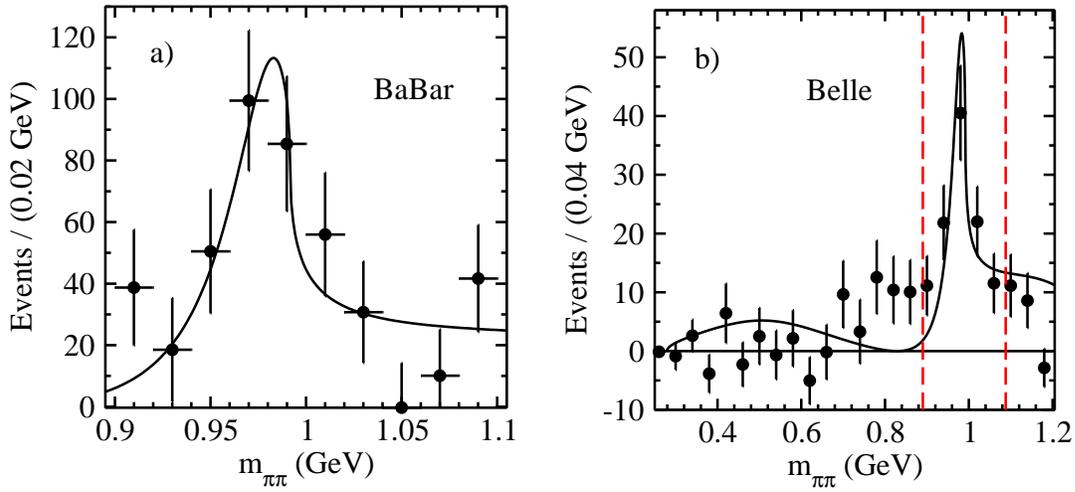

\subfigure{\includegraphics[width=0.45\textwidth]{babarpm.eps}}~~~~~
\subfigure{\includegraphics[width=0.45\textwidth]{belle0.eps}}
\caption{$\pi^+\pi^-$ mass distributions in the 
$B^\pm\to \pi^+\pi^-K^\pm$ (a) and in the  
$B^0 \to \pi^+\pi^-K^0$ decays (b). Vertical lines delimit a band of the 
$f_0(980)$ events. }
\end{figure}


\begin{thebibliography}{99}
\bibitem {baba03}
B.~Aubert \textsl{et al.}, BaBar Collaboration, 
Phys. Rev. \textbf{D70} (2004) 092001.
\bibitem {baba04}
B.~Aubert \textsl{et al.}, BaBar Collaboration, 
Phys. Rev. Lett. {\bf 94} (2005) 041802.
\bibitem {bell0412}
A.~Garmash \textsl{et al.}, Belle Collaboration, 
Phys. Rev. \textbf{D71} (2005) 092003.
\bibitem {chen04} 
K.-F.~Chen\textsl{ et al.}, (Belle Collaboration), Phys. Rev. \textbf{D72} (2005)
 012004.
\bibitem {kami97}
R.~Kami\'nski, L.~Le\'sniak and B. Loiseau, 
Phys. Lett. \textbf{B413} (1997) 130.
\bibitem {groo03}
N.~de~Groot, W.~N.~Cottingham and I.~B.~Whittingham, 
Phys. Rev. \textbf{D68} (2003) 113005. 
\bibitem {ciuc04}
M.~Ciuchini, E.~Franco, G.~Martinelli, A.~Masiero, M.~Pierini and 
L.~Silvestrini, talk given at 
39th Rencontre de Moriond
, La Thuile, Aosta Valley, Italy, 21st-28th March 2004, 
hep-ph/0407073.
\bibitem {FKLL}
A.~Furman, R.~Kami\'nski, L.~Le\'sniak and B. Loiseau, 
Phys. Lett. \textbf{B622} (2005) 207.
\bibitem {meis01}
U.-G.~Mei\ss ner and J~.~A.~Oller, Nucl. Phys. \textbf{A679} (2001) 671.

\end{thebibliography}
\end{document}